\documentclass[twocolumn]{aastex63}

\usepackage{graphicx}

\begin{document}

\title{Measuring Phase Errors in the Presence of Scintillation} 

\author{Justin R. Crepp, Stanimir O. Letchev, Sam J. Potier, Joshua H. Follansbee, Nicholas T. Tusay}
\affiliation{Department of Physics University of Notre Dame, 225 Nieuwland Science Hall, Notre Dame, IN 46556, USA}

\correspondingauthor{Justin R. Crepp}
\email{jcrepp@nd.edu}

\begin{abstract}
    Strong turbulence conditions create amplitude aberrations through the effects of near-field diffraction. When integrated over long optical path lengths, amplitude aberrations (seen as scintillation) can nullify local areas in the recorded image of a coherent beam, complicating the wavefront reconstruction process. To estimate phase aberrations experienced by a telescope beam control system in the presence of strong turbulence, the wavefront sensor (WFS) of an adaptive optics must be robust to scintillation. We have designed and built a WFS, which we refer to as a ``Fresnel sensor,'' that uses near-field diffraction to measure phase errors under moderate to strong turbulent conditions. Systematic studies of its sensitivity were performed with laboratory experiments using a point source beacon. The results were then compared to a Shack-Hartmann WFS (SHWFS). When the SHWFS experiences irradiance fade in the presence of moderate turbulence, the Fresnel WFS continues to routinely extract phase information. For a scintillation index of $S = 0.55$, we show that the Fresnel WFS offers a factor of $9\times$ gain in sensitivity over the SHWFS. We find that the Fresnel WFS is capable of operating with extremely low light levels, corresponding to a signal-to-noise ratio of only $\mbox{SNR}\approx 2-3$ per pixel. Such a device is well-suited for coherent beam propagation, laser communications, remote sensing, and applications involving long optical path-lengths, site-lines along the horizon, and faint signals.
    \vspace{0.3in}
\end{abstract}

\section{Introduction}\label{sec:intro}

As a coherent beam of light passes through a turbulent medium, near-field diffraction effects transform phase variations into amplitude variations (and vice-versa). The resulting constructive and destructive interference pattern---known as scintillation---can limit the ability of an adaptive optics (AO) system to sense changes in the wavefront. Scintillation occurs when operating under strong turbulent conditions (high Rytov number) or over long optical path lengths. The challenge of dealing with diffractive amplitude variations is often referred to as ``deep turbulence'' \citep{watnik_2018}.

In the case of the Shack-Hartmann wavefront sensor (SHWFS), mild-to-moderate amplitude aberrations cause scintillation bias. The centroid location of a focal plane spot is set by the weighted-average of the intensity across a subaperture (``c-tilt''), whereas the reconstructor assumes that spot locations correspond to the geometric average or gradient slope (``g-tilt'') across a subaperture \citep{barchers_2002}. As amplitude aberrations increase from moderate levels to strong levels, the intensity of imaged spots diminishes in areas of the beam that experience destructive interference. The resulting irradiance fade makes local slope measurements highly uncertain. When a large fraction of focal plane spots disappear, the entire reconstruction process may become unreliable \citep{dubose_2020}. Irradiance fade, branch points, and branch cut phase discontinuities motivate the need to explore alternative sensing technologies \citep{spencer_2015,spencer_2017}.

\begin{figure*}[t]
    \centering
    \includegraphics[width=\textwidth]{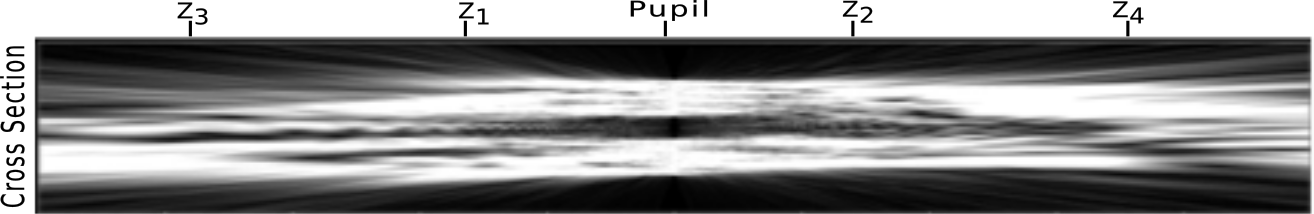}
    \caption{Beam cross-section showing light propagating from left to right through a telescope. Diffraction caused by aberrations creates measurable changes in intensity along the optical axis. The nlcWFS records images at locations $z=\{z_1, z_2, z_3, z_4\}$. A physical optics model retrieves phase and amplitude at the reimaged pupil plane. Figure adapted from \citealt{crass_2012}.}
    \label{fig:beam}
\end{figure*}

The phase and amplitude of a propagating beam of light may be retrieved by measuring the intensity of diffracted light at several locations along the optical axis (Fig.~\ref{fig:beam}). Originally developed for applications in astronomy, a WFS that relies on multiple camera images (generally three or more) recorded on either side of the optical system pupil would offer a large capture range and excellent sensitivity \citep{guyon_2010}. Such a device requires an order of magnitude fewer photons than the SHWFS to reach the same level of wave-front error \citep{crass_2014,mateen_2015}. As a consequence of the macroscopic separations between detector planes---each located outside of the pupil plane---sensor performance is also predicted to exceed that of a curvature sensor by several orders of magnitude \citep{guyon_2010}. Initially named a ``non-linear curvature WFS'' in the literature, we refer to the device as a ``Fresnel WFS'' due to the physical optics methods needed to reconstruct the complex field \citep{guyon_2010}.\footnote{The multiple camera images have been described as being ``out of focus'' or purposefully defocused including displacements many times larger than the beam diameter.} 

In this paper, we show that the Fresnel WFS (hereafter FWFS) can accommodate amplitude variations by measuring near-field diffraction effects along the propagation path. In this regard, scintillation may be used to infer wavefront information rather than acting as a noise source---allowing AO systems to operate under a wide range of atmospheric conditions. We have designed and built a prototype FWFS at the University of Notre Dame. We describe laboratory experiments that demonstrate reconstruction precision and accuracy relevant to modern AO systems. We also compare the sensitivity of the FWFS to a SHWFS as a function of incident flux level to study performance in photon-noise-limited and read-noise limited measurement regimes. 

\section{Experimental Methods}\label{sec:experiment}


\subsection{Experimental Design}\label{sec:design}

Turbulence sensing experiments were conducted in the Beam Control Lab at the University of Notre Dame. Primary objectives of the experiment were to validate reconstruction algorithms using hardware and to quantify performance of the FWFS versus the SHWFS in the presence of moderate scintillation. Three different experiments were conducted to study the FWFS.

In the first experiment, a custom aberrator was fabricated by etching the letters ``ND'' (short for ``Notre Dame'') into a glass substrate using photolithography to test the FWFS reconstructor. In the second experiment, a randomized pattern more representative of atmospheric turbulence was developed by spraying acrylic onto a transparent substrate. The resulting aberrations were measured and compared to that of a SHWFS. In the third experiment, the same (static, non-rotating) phase plate was used to study the sensitivity of each sensor in the limit of diminishing flux levels. 

\subsection{Component Lay-out}

All measurements were conducted with monochromatic, coherent light ($\lambda=532$ nm). The optical system was illuminated using a laser-diode-pumped DPSS Laser Module with 0.9 mW power (Thorlabs CPS532-C2). Light was injected into a single mode fiber to create a point source (Thorlabs P1-405B-FC-1). After collimation, an optical relay was used to establish the initial beam diameter. Aberrations were introduced by inserting a transmissive phase plate in front of the optical system entrance pupil which had a diameter of 0.7 mm. The system entrance pupil and pseudo-telescope was defined using a pin-hole to create a circular diffraction pattern. 

\begin{figure}[t]
    \centering
    \includegraphics[width=0.47\textwidth]{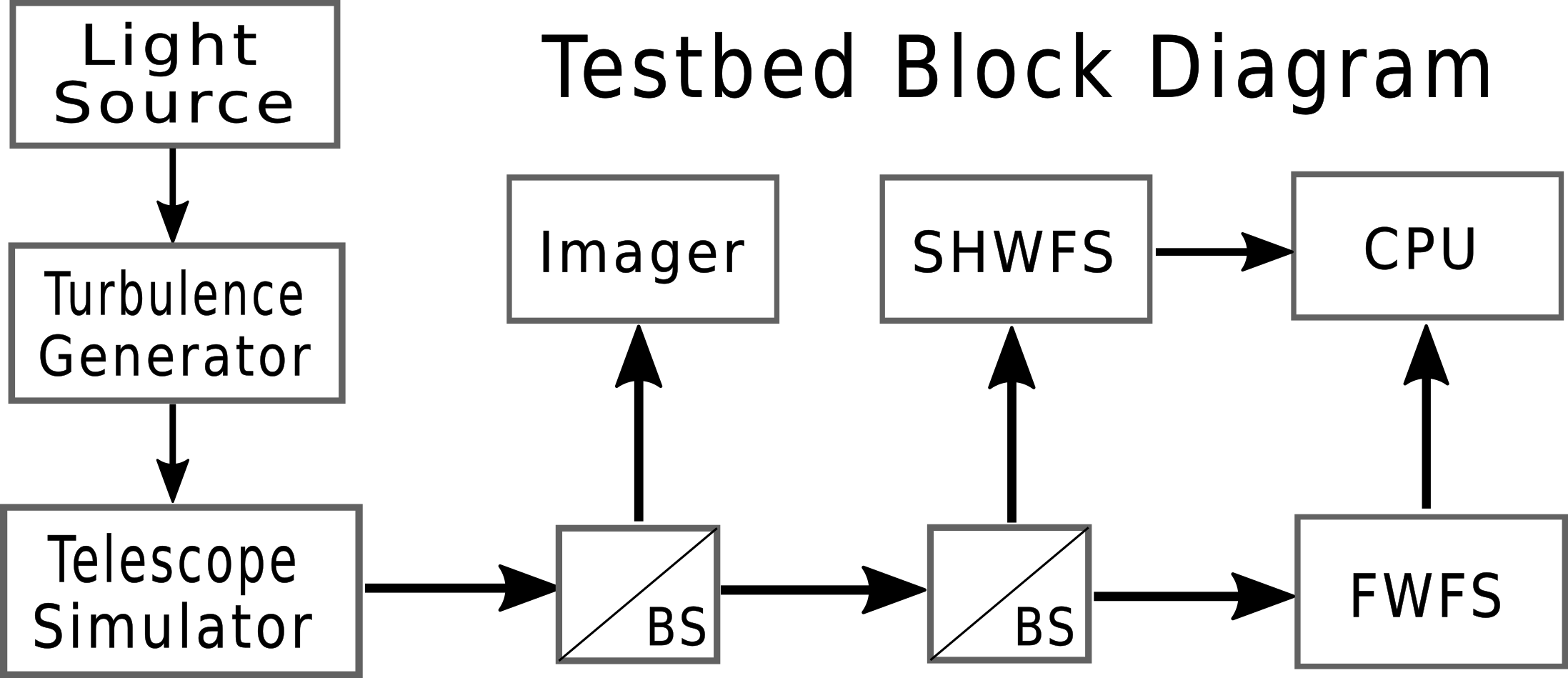}
    \caption{Simplified schematic describing the experiment.}
    \label{fig:block}
\end{figure}

Downstream components comprising the AO sensing modules included a SHWFS (WFS30-7AR) and custom-built, four-plane FWFS. The SHWFS used $19\times19$ sub-apertures with 150 $\mu$m lenslet pitch. In the focal plane, the SHWFS sampled the FWHM of lenslet spots using $3.2\pm0.1$ pixels. Approximately 26 pixels separated individual spots in each direction to prevent cross-talk. The FWFS used an Andor Zyla 4.2 sCMOS camera with 6.5 $\mu$m pixel pitch. Pixel sampling was selected based on optical constraints for beam diameter and in an effort to compare results to the $19\times19$ SHWFS. Selection of Fresnel plane locations is discussed in $\S$\ref{sec:reconstruction}. High-quality beam-splitters created separate wavefront sensing channels to allow contemporaneous measurements using both SHWFS and FWFS devices. Figure~\ref{fig:block} shows a simplified block diagram of the experimental lay-out. 


\subsection{Fabrication of the ``ND'' Mask}

A custom ``ND'' mask was fabricated using photolithography at the Notre Dame Nanoscience and Technology lab. First, a thin layer of chromium (100-200 nm) was deposited onto a 5 mm thick fused-silica glass substrate. Next, a 1-2 micron layer of photoresist was deposited on top of the chromium. A custom mask was applied to the photoresist layer and ultra-violet light ($\lambda_{\rm UV}=365-405$ nm) used to imprint the letters ``ND.'' The mask was removed and the pattern in the photoresist layer eroded chemically. Dry plasma etching was used to create pseudo-square walls. Finally, the chrome layer was removed with a Cerium IV nitrate solution.  

\begin{figure}[t]
    \centering
    \includegraphics[width=0.26\textwidth,trim={0 0mm 0 0},clip]{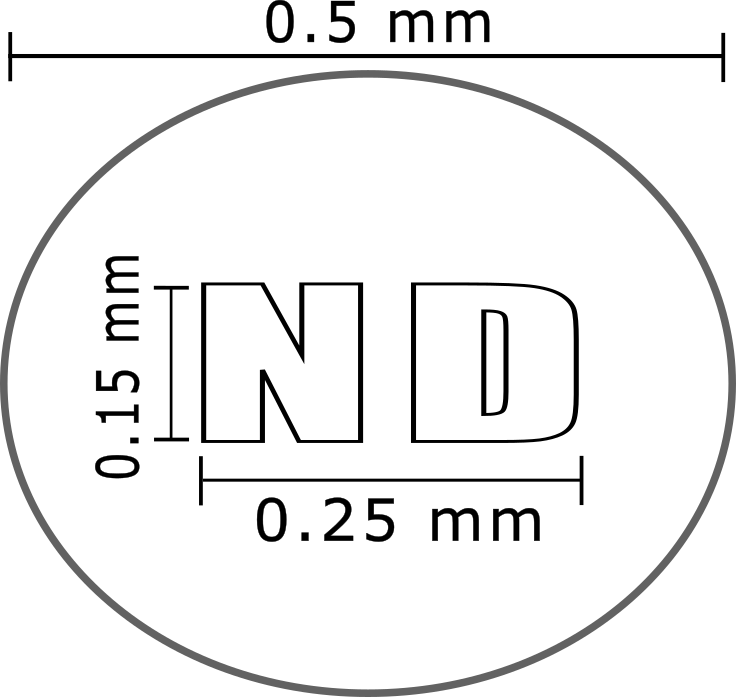}
    \caption{Photolighography ``ND'' mask design.}
    \label{fig:mask_design}
\end{figure}

The ``ND'' letters were designed to a height of 150 $\mu$m and 250 $\mu$m in collective width. Figure~\ref{fig:mask_design} shows a diagram of the design with physical dimensions. The outer circle diameter represents the approximate beam illumination area. The goal fabrication depth was -100 nm. An Olympus (\#OLS4100) profilometer was used to measure the depth of the final etched pattern. Reconstruction results of the ND mask are shown in $\S$\ref{sec:nd}.

\subsection{Sensor Comparison}\label{sec:sensors}

The second experiment used a plastic substrate sprayed with acrylic to generate aberrations that better mimic atmospheric turbulence. The non-rotating substrate was held in a static position located approximately 1.5 cm in front of the telescope entrance aperture. Both sensors were used to measure the same phase disturbance. Reconstruction results for the SHWFS and FWFS are compared in $\S$\ref{sec:comparison}.

The third experiment was designed to assess wavefront error residuals as a function of incident flux. Flux levels were varied from high signal-to-noise ratio (SNR) intensity measurements to low SNR intensity measurements. The amount of light was incrementally decreased by adjusting fiber coupling efficiency and integration time. Flux levels were reduced by several orders of magnitude overall until each sensor could no longer reconstruct the wavefront. Equitable comparison between sensors was made by calibrating the number of photo-electrons detected in each channel and correcting for detector gain. The location of the phase plate was also moved further from the telescope entrance aperture to induce stronger amplitude aberrations (scintillation) ($\S$\ref{sec:amplitude}). 


\subsection{Wavefront Reconstruction}\label{sec:reconstruction}

The SHWFS was used for the second and third experiments ($\S$\ref{sec:design}). The collimated beam diameter at the SHWFS was magnified to fit across the $19\times19$ subaperture array. Several commercial, custom, and publicly-available reconstruction algorithms were tested to register, calibrate, and analyze the SHWFS measurements. Included in the analysis are three independent methods: Thorlabs commercial software; a custom Zernike polynomial fitting program written by our team in Matlab; and an open-source Zernike modal method downloaded from Github \citep{antonello_2014}.\footnote{\url{https://github.com/jacopoantonello/mshwfs}} Our custom reconstruction algorithm, which performed similar calculations to the Thorlabs software, was written to by-pass warning messages of the commercial program when operating at the lowest flux levels. The publicly available method, which ``computes definite integrals of the gradients of the Zernike modes within each subaperture,'' was also written in Matlab and used in \citealt{lechner_2019} to study scintillation.

The number of reconstructed FWFS spatial samples across each dimension of the ``ND'' letters spanned approximately $35 \times 60$ pixels after accounting for beam magnification. Due to diffraction, the number of pixels illuminated on the camera depends on the z-distance of each observing plane. In the first experiment, a single camera with 4.8 $\mu$m pitch was translated along the optical axis to acquire images at z-distances of $z_1=-1$ cm, $z_2=+1$ cm, $z_3=-2$ cm, and $z_4=+2$ cm from the reimaged pupil (see Fig.~\ref{fig:beam}). Such a geometry enables sensitivity to high spatial frequency aberrations. 

In the second and third experiments, all four ``defocused'' images were placed onto the Andor camera for the Fresnel sensor. Pixel sampling was selected based on optical constraints for beam diameter and in an effort to compare results to the SHWFS. A native sampling (defined at the pupil for $z=0$) of $107\times107$ pixels was used across each Fresnel image. Results were then binned in software to $19\times19$ across the pupil to match the SHWFS lenslets. Fresnel plane locations were selected by simulating various $z$-distances given a Kolmogorov-like spectrum. Simulation results were used as a starting point for the experiment and plane locations were then adjusted to optimize performance. The final configuration used distances of $z_1=-2$ cm, $z_2=+2$ cm, $z_3=-9$ cm, and $z_4=+9$ cm, a geometry that enables sensitivity to a broader range of spatial frequencies compares to the first experiment.

The FWFS used a custom reconstruction method developed in MATLAB. The program is based on a modified version of the Gerchberg-Saxton algorithm that uses four planes \citep{gerchberg_1972}. A Fourier transform-based phase-unwrapping algorithm was developed based on \citealt{schofield_2003}. Both programs were converted to C++ to facilitate laboratory signal processing and communications and to prepare for closed-loop experiments.

\section{Results}\label{sec:results}

\subsection{Reconstruction of the ``ND'' Mask}\label{sec:nd}

\begin{figure}[t]
    \centering
    \includegraphics[width=0.52\textwidth,trim={19mm 21mm 0 0},clip]{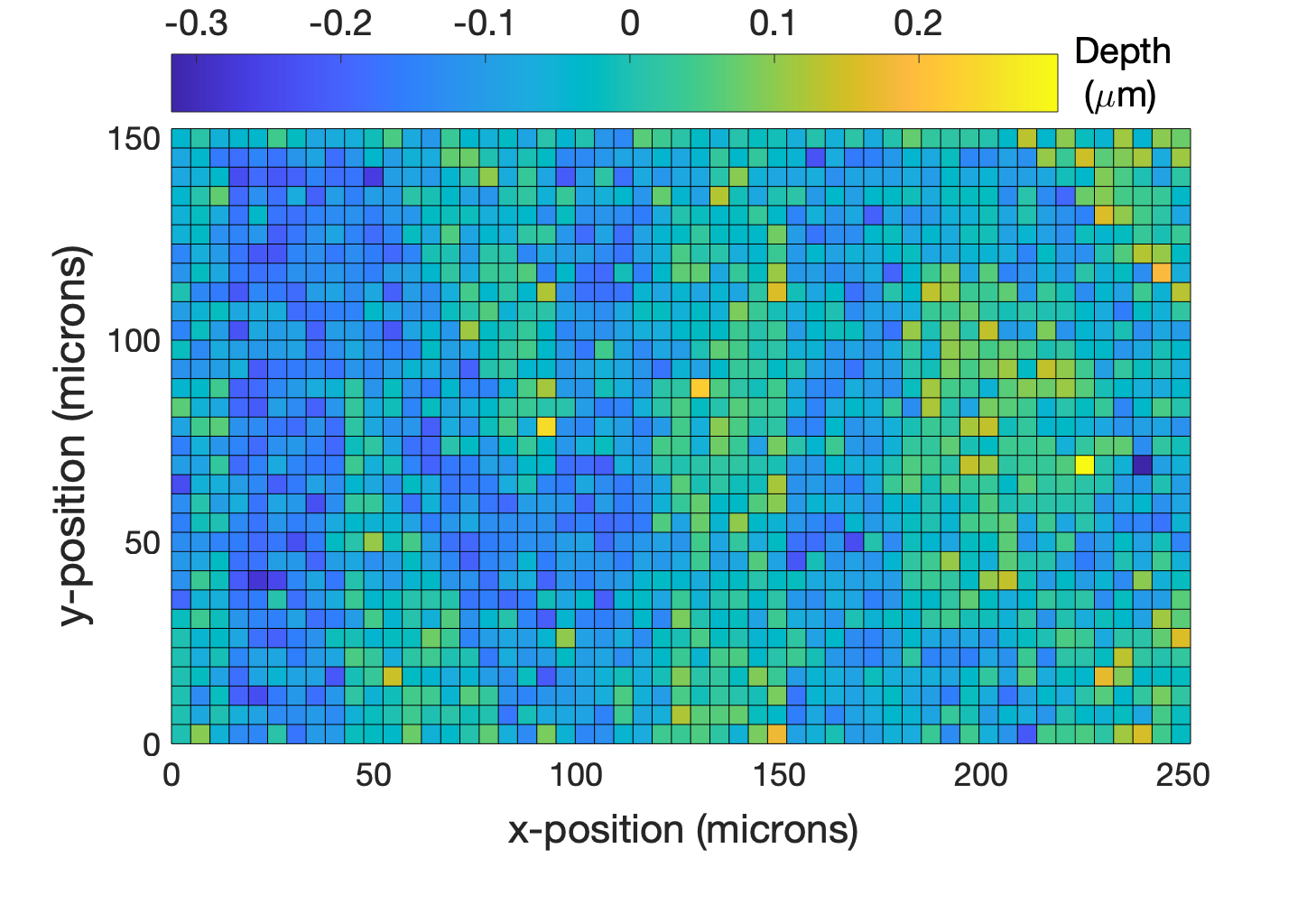}
    \caption{Profilometer measurements of the -100 nm depth ``ND'' photo-lithography mask.}
    \label{fig:profilometer}
\end{figure}

\begin{figure}[t]
    \centering
    \includegraphics[width=0.52\textwidth,trim={3mm 0mm 3mm 1mm},clip]{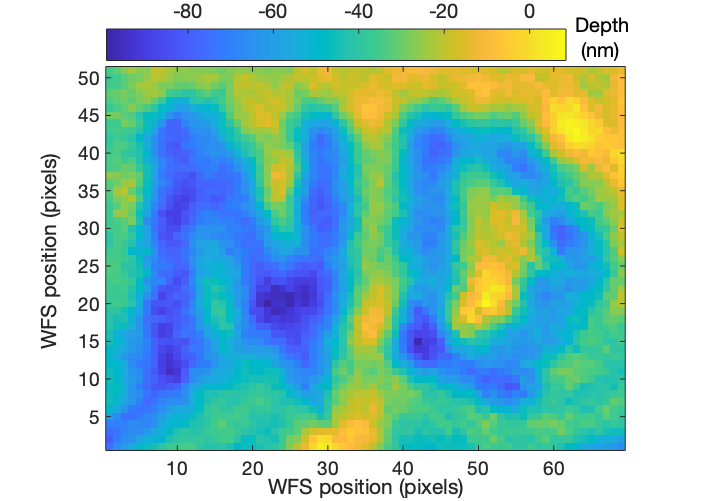}
    \caption{Phase reconstruction of the -100 nm depth ``ND'' photo-lithography mask using the FWFS.}
    \label{fig:ndmask_reconstruction}
\end{figure}

Fig.~\ref{fig:profilometer} shows results for profilometer measurements of the ``ND'' mask. The depth was found to be $-113 \pm 38$ nm. Figure~\ref{fig:ndmask_reconstruction} shows wavefront reconstruction results using FWFS measurements. The ``ND'' letters are easily recovered, although diffraction appears to distort the pattern. The average depth of the ``ND'' mask was found to be $-78 \pm 6$ nm using the FWFS. Features of the mask---individual letter boundaries---were presumed to be too sharp (very high spatial frequency) for the SHWFS to reconstruct, so it was not used for this particular experiment. 

Comparing Fig.~\ref{fig:profilometer} with Fig.~\ref{fig:ndmask_reconstruction} and the scale of the depth measurements (microns versus nanometers), the FWFS phase results display less scatter than the profilometer (better precision). While questions remain regarding the absolute accuracy of both devices, the measured depths are consistent with one another to within the uncertainties ($1 \sigma$). This experiment was the first to validate the FWFS hardware and reconstruction algorithm using known shapes. 

\begin{figure*}
    \centering
    \includegraphics[width=0.48\textwidth]{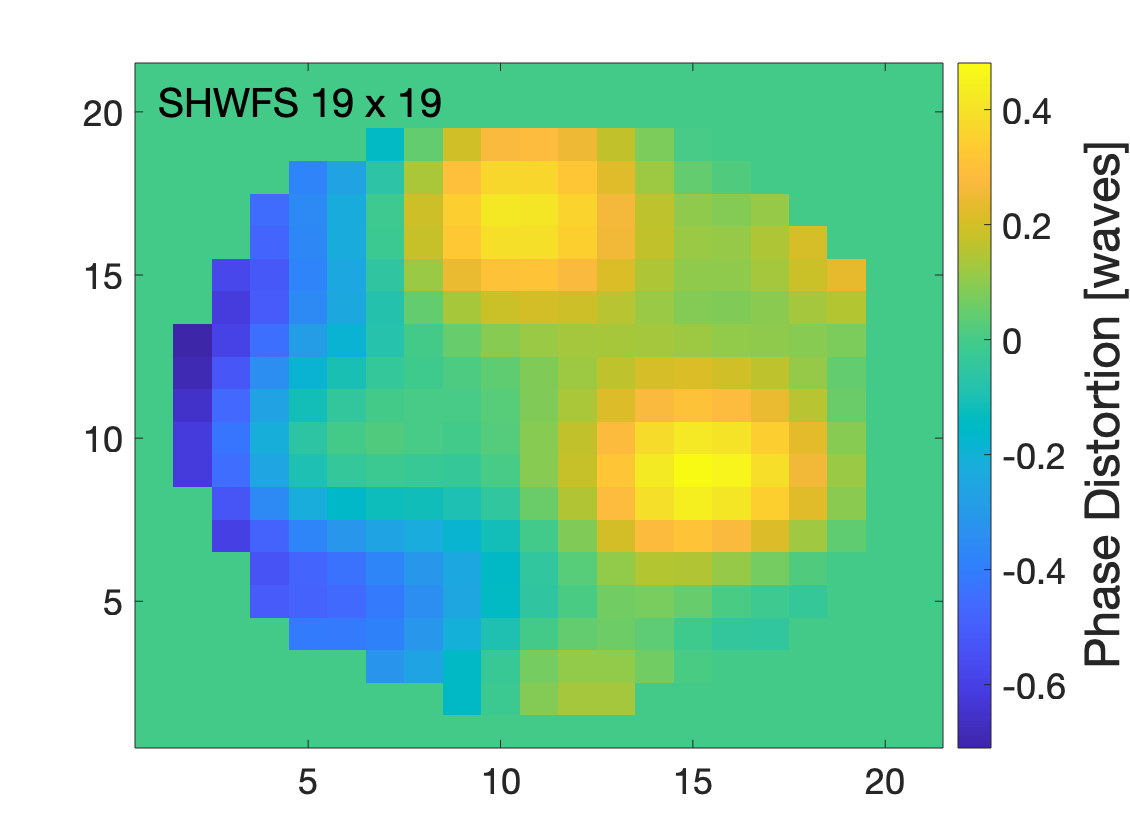}
    \includegraphics[width=0.48\textwidth]{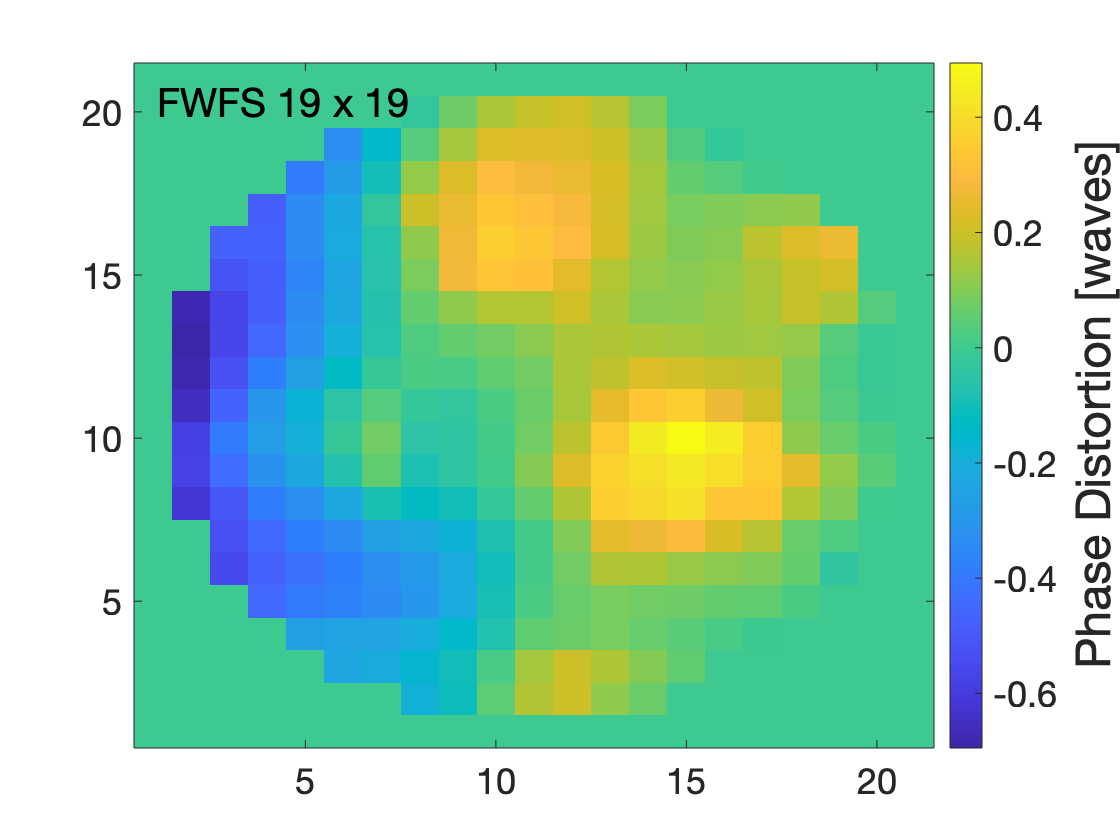}
    \caption{Reconstruction of randomized pattern caused by spraying acrylic onto a glass disk with the (left) SHWFS and (right) FWFS. Results are equivalent to within 0.09 waves rms.}
    \label{fig:acrylic}
\end{figure*}

 \begin{figure*}[t]
    \centering
    \includegraphics[width=0.48\textwidth,trim={16mm 26mm 3mm 1mm},clip]{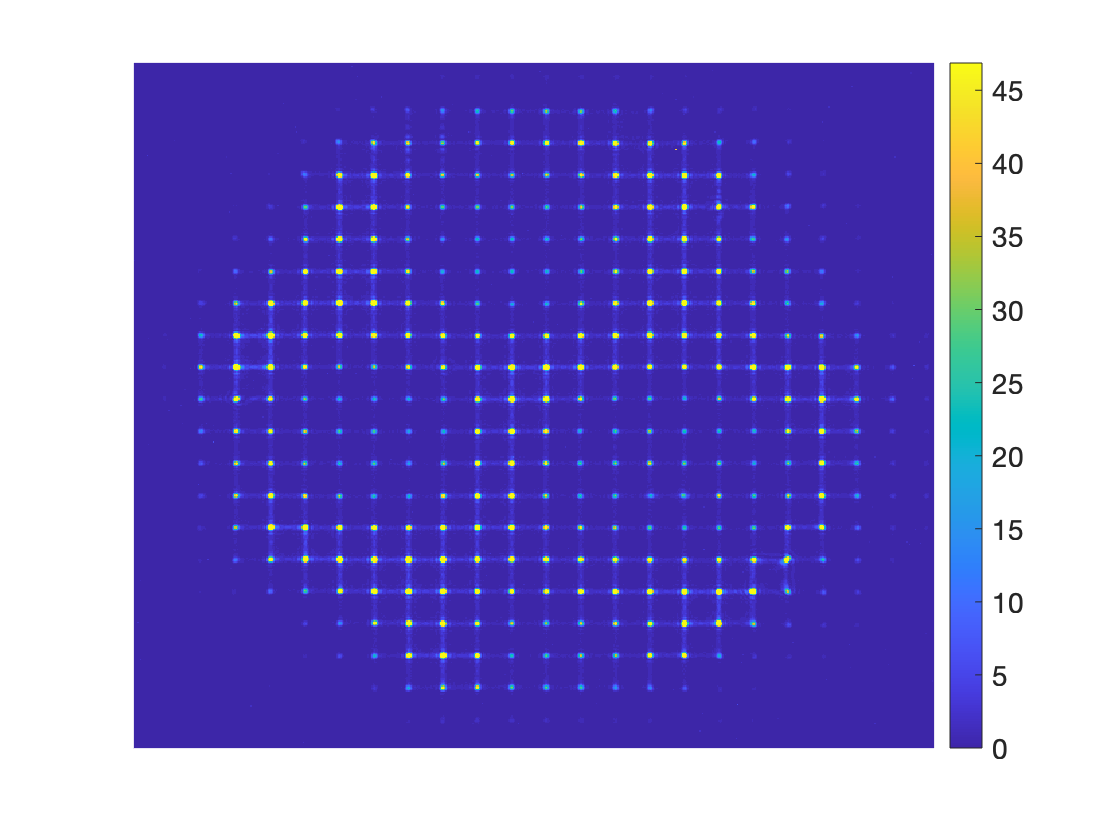}
    \hspace{5mm}
    \includegraphics[width=0.48\textwidth,trim={16mm 26mm 3mm 1mm},clip]{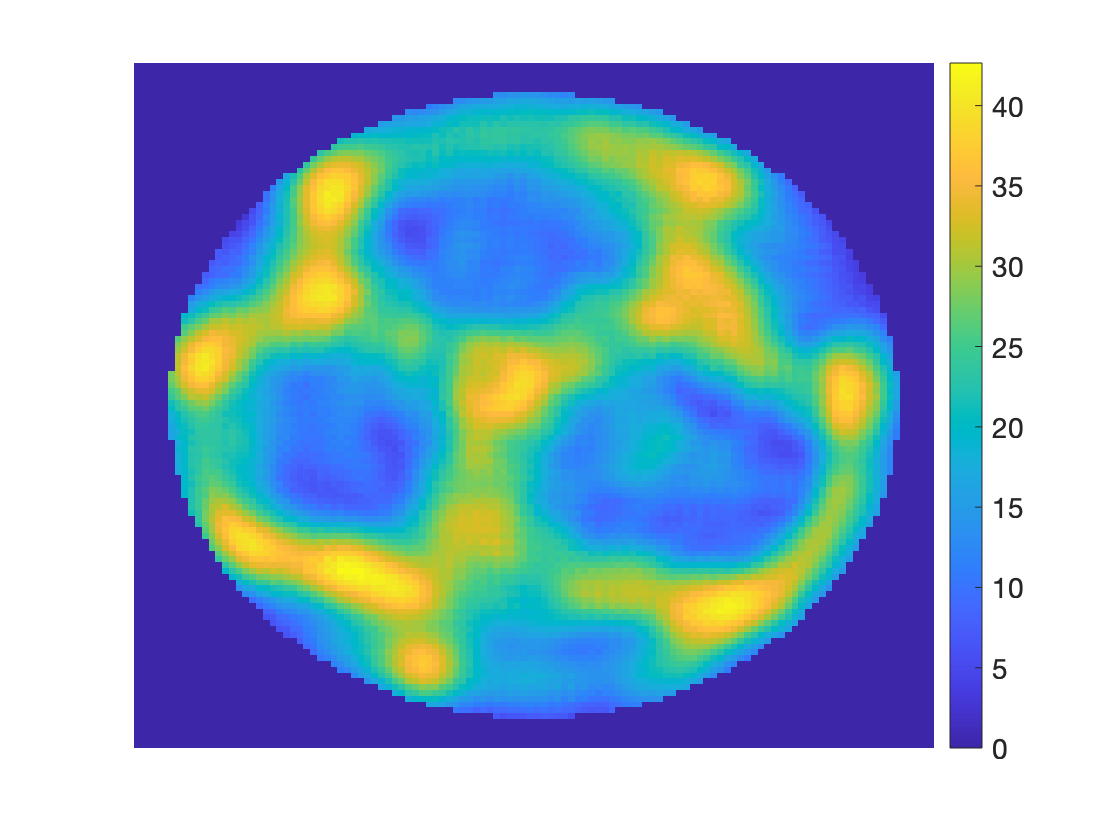}
    \caption{(left) SHWFS amplitude variations showing the effects of irradiance fade from near-field diffraction. (right) Amplitude variation map reconstructed by the FWFS at $z=0$. Results between sensors are highly correlated.}
    \label{fig:amp_comparison}
\end{figure*}

\subsection{Aberrated Beam Phase Reconstruction}\label{sec:comparison}

The second experiment introduced lower spatial frequency aberrations and more representative of atmospheric turbulence compared to the ``ND'' mask. Figure~\ref{fig:acrylic} shows wavefront reconstruction results for the acrylic phase plate using both the SHWFS and FWFS. Pixels have been interpolated in software and binned to ensure equivalent spatial sampling with each sensor (19 $\times$ 19). We find that the SHWFS and FWFS produce nearly identical phase reconstructions. The rms wavefront difference between devices was found to be 0.09 waves (48 nm) after removing an offset in piston. Several different samples through the acrylic plate were studied, yielding similar results. 


\subsection{Amplitude Aberrations}\label{sec:amplitude}

In the third experiment, the distance of the turbulence generator from the entrance pupil was adjusted to increase the amount of amplitude aberrations induced through near-field diffraction. Figure~\ref{fig:amp_comparison} displays results for the beam amplitude. In the SHWFS channel, the amplitude is estimated using spot field intensities. In the FWFS channel, amplitude aberrations at the reimaged pupil ($z=0$) are reconstructed using phase diversity provided by the four out-of-pupil plane images. 

The SHWFS spot field illustrates how scintillation causes local irradiance fade across portions of the beam. As discussed below, irradiance fade ultimately limits the ability of the SHWFS to reconstruct the wavefront when dealing with faint sources. The fact that the FWFS uses (constructive and) destructive interference to inform the reconstruction process suggests that the device may be used in the presence of strong scintillation provided that diffracted light is captured by the camera. 

The amplitude patterns measured by each sensor are highly correlated since they are derived from the same beam and aberration source, differing only by a small amount of non-common-path errors. Using the technique of \citealt{andrews_2001}, we measure the normalized variance in intensity fluctuations. A scintillation index less than unity may be considered weak or moderate turbulence, whereas larger $S$ values may be considered ``deep turbulence.'' We estimate a scintillation index of 
\begin{equation}
    S=\frac{<I^2> - <I>^2}{<I>^2}=0.55
\end{equation}
for the third experiment indicating moderate turbulence using a single phase plate. Assuming Kolmogorov-like turbulence, the measured scintillation index corresponds (roughly) to an isoplanatic angle of $\theta_0=0.25$ arcseconds and $r_0=1.4$ cm at zero zenith angle and average turbulence height of $h=3.5$ km based on data from \citealt{sarazin_02}. The beam profile shown in Fig.~\ref{fig:acrylic} and Fig.~\ref{fig:amp_comparison} was then used to study the sensitivity limit of each sensor. 

\begin{figure*}[t]
    \centering
    \includegraphics[width=0.9\textwidth]{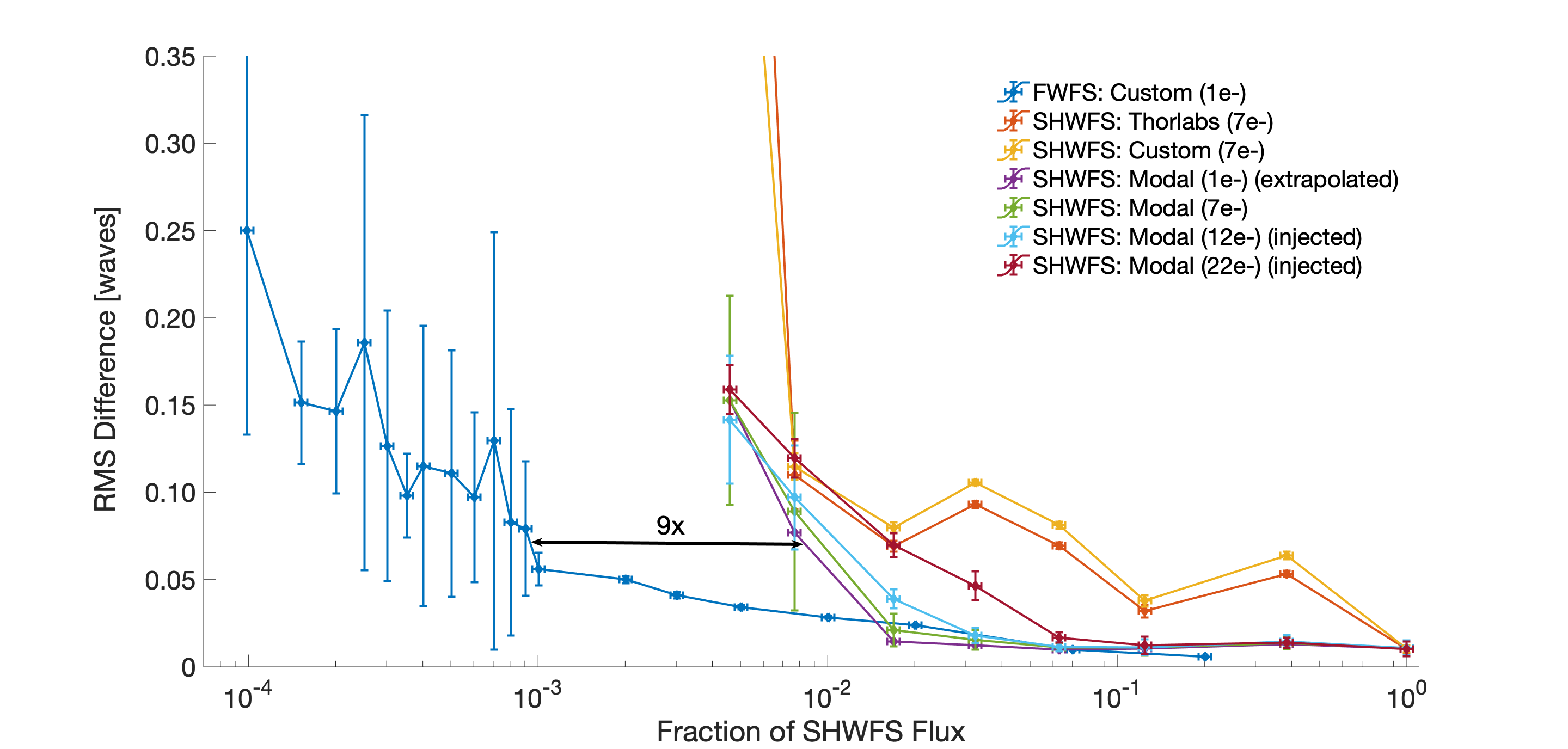}
    \caption{Sensitivity comparison between SHWFS and FWFS. The FWFS offers a factor of 9x improvement in the ability to operate with lower flux levels for a scintillation index of $S=0.55$. See text for discussion.}
    \label{fig:sensitivity}
\end{figure*}

\subsection{Sensitivity Comparison}\label{sec:sensitivity}

After demonstrating that the FWFS accurately reconstructs wavefront phase and amplitude, we then adjusted the amount of light entering each sensor. Photon-noise-limited performance was studied by systematically decreasing input laser intensity to the experiment. Residual wavefront error was calculated for each incident flux level to assess performance. The reconstructed wavefront phase was compared to measurements with the highest signal-to-noise ratio (SNR). Each sensor served as its own calibrator. Low SNR SHWFS measurements were compared to the highest SNR SHWFS measurement, and low SNR FWFS measurements were compared to the highest SNR FWFS measurement.

To generate a comparison plot, signal levels were normalized based on incidence flux measured in the SHWFS channel when using the brightest laser intensity. Decreasing the integration time of each sensor and changing fiber coupling efficiency into the upstream beam allowed the experiment to span four orders of magnitude in dynamic range. Reconstruction results for 15 sequential images were averaged together at each flux level for each sensor. The standard deviation between independent measurements and the reference wavefront is taken as the uncertainty in reconstructed wavefront error for each data set (represented by vertical error bars in the sensitivity plots). 

\subsubsection{Sensitivity Limits}\label{sec:limits}

Significant care was taken to calibrate each sensor by taking into account integration time and detector gain values. While each sensor has a different quantum efficiency, $\approx75$\% for the SHWFS (Thorlabs monochrome DCC3260M) and $\approx80$\% for the FWFS (Andor Zyla 4.2 Plus),\footnote{Vendor QE specifications evaluated at $\lambda=532$ nm.} summing the number of detected photo-electrons and normalizing the results accounts for systematic differences between sensing channels. To further ensure that the calibration process was accurate, we also temporarily placed the SHWFS device into the FWFS channel to measure the relative flux delivered by the beam-splitter to each path.

Figure~\ref{fig:sensitivity} shows residual wavefront phase error as a function of incident flux measured relative to the brightest SHWFS data set. Wavefront error increases with decreasing flux level for all reconstruction methods, allowing for the relative performance of each sensor to be compared. As with the first and second experiment, systematic differences in reconstructed phase between the two sensors is negligibly small for the high SNR measurements. Considering the well-calibrated commercial SHWFS device as a truth sensor, we find that the accuracy of the FWFS is better than $\lambda/10$ rms. 

The various SHWFS reconstruction results (see $\S$\ref{sec:reconstruction} for details of reconstruction methods) are shown alongside the FWFS data (Fig.~\ref{fig:sensitivity}). The default SHWFS commercial software (``Thorlabs'') and our in-house reconstructor (``Custom'') perform comparably, asymptoting just below 1\% of the brightest incident laser intensity for the experiment. The publicly-available SHWFS reconstructor (``Modal'') offers the best performance at all flux levels, allowing for approximately 1.7$\times$ fainter signals to be reconstructed prior to failing. 

We find that the FWFS is able to reconstruct wavefront phase aberrations at flux levels approximately $9\times$ lower than that of the SHWFS before asymptoting. Given the consistency in results for the various SHWFS reconstruction algorithms, it was concluded that software was not the limiting factor for the SHWFS measurements. We note that at flux values of $\approx 2 \times10^{-2}$ the SHWFS appears to perform slightly better than the FWFS. However, these data points are either extrapolated to lower read noise or are consistent to within 1 sigma. Further, the scale of the difference in aberrations at this level is only 0.02 waves, which is below the calibrated accuracy of either sensor in this experiment. To more fully understand the limits of each sensor, we next studied the impact of read noise on the sensitivity curves.  

\begin{figure*}[t]
    \centering
    \includegraphics[width=0.9\textwidth]{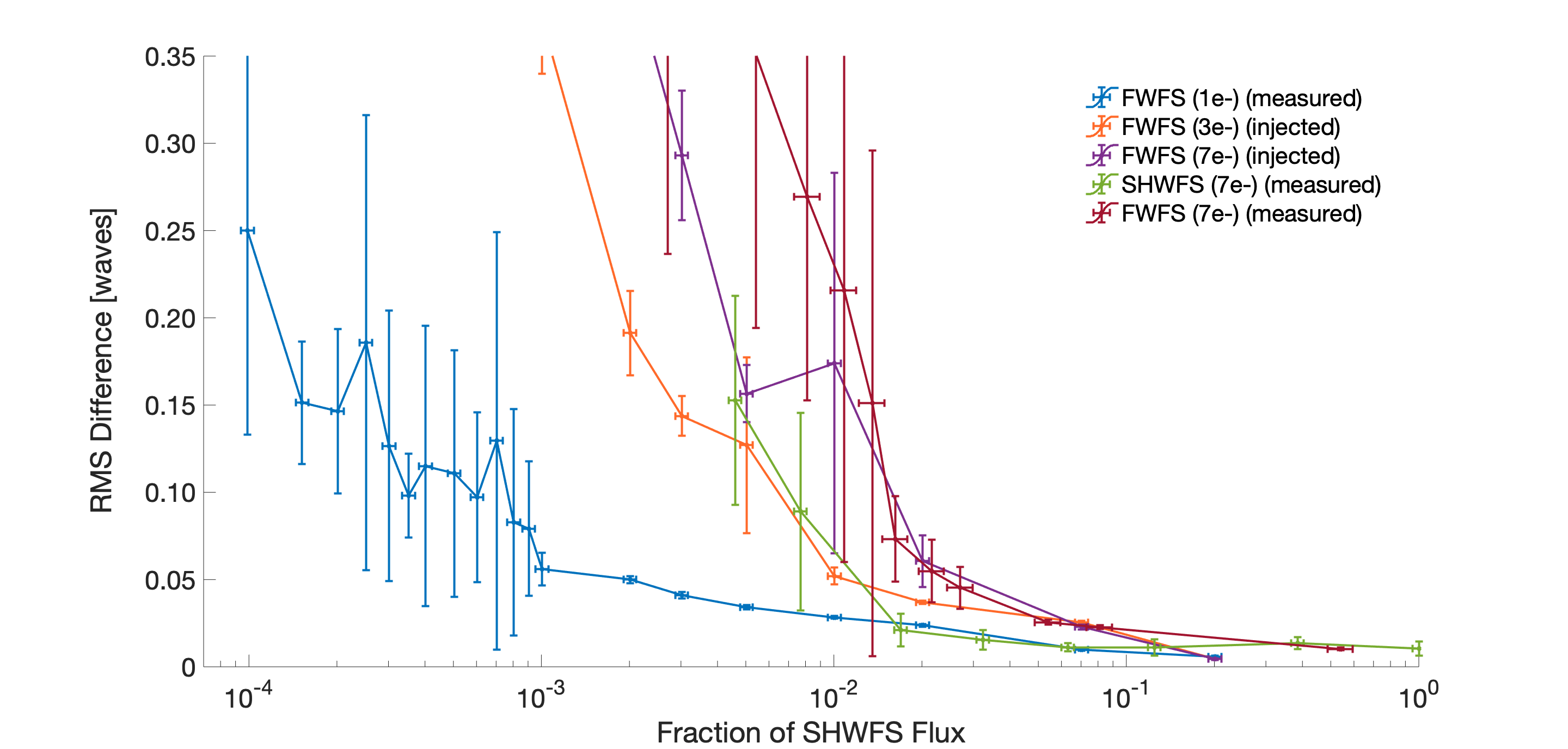}
    \caption{Experimental results for the FWFS as a function of read noise, both injected using Andor camera darks as well as with a physically different detector. The $7e-$ read noise SHWFS data set is included for comparison. See text for discussion.}
    \label{fig:FWFS_RN}
\end{figure*}

\begin{figure*}[t]
    \centering
    \includegraphics[width=0.46\textwidth]{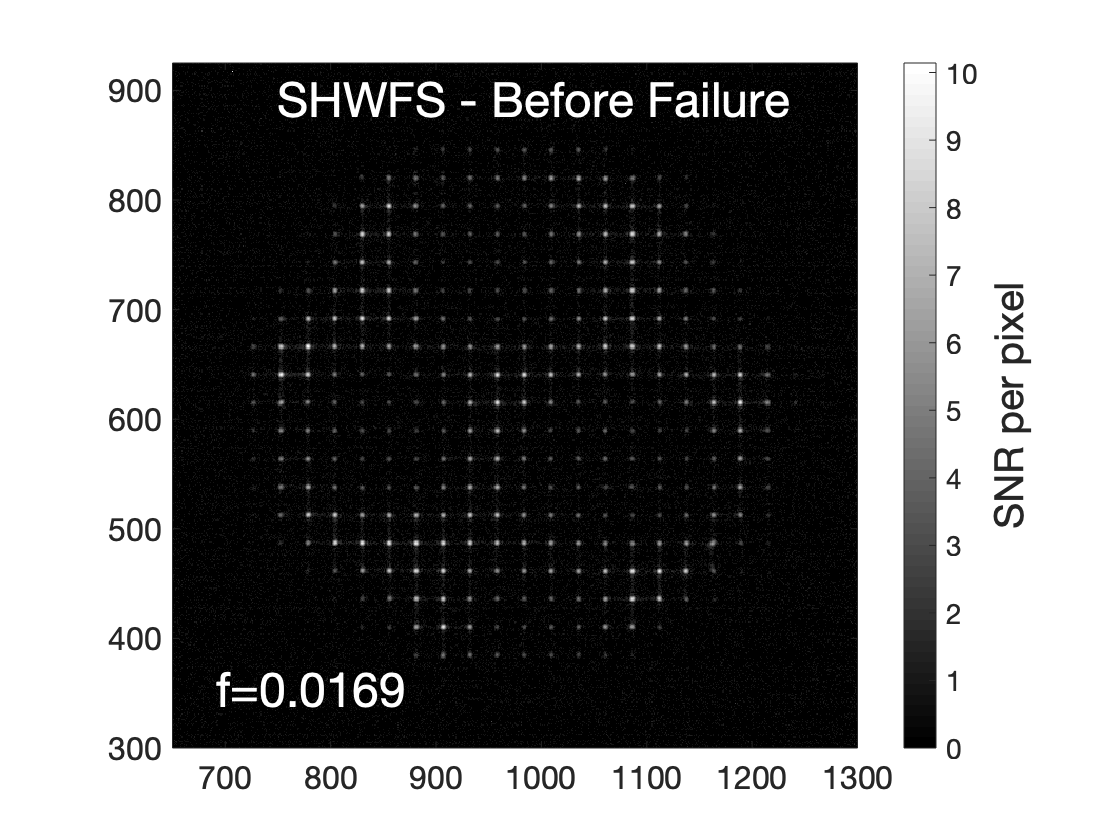}
    \includegraphics[width=0.46\textwidth]{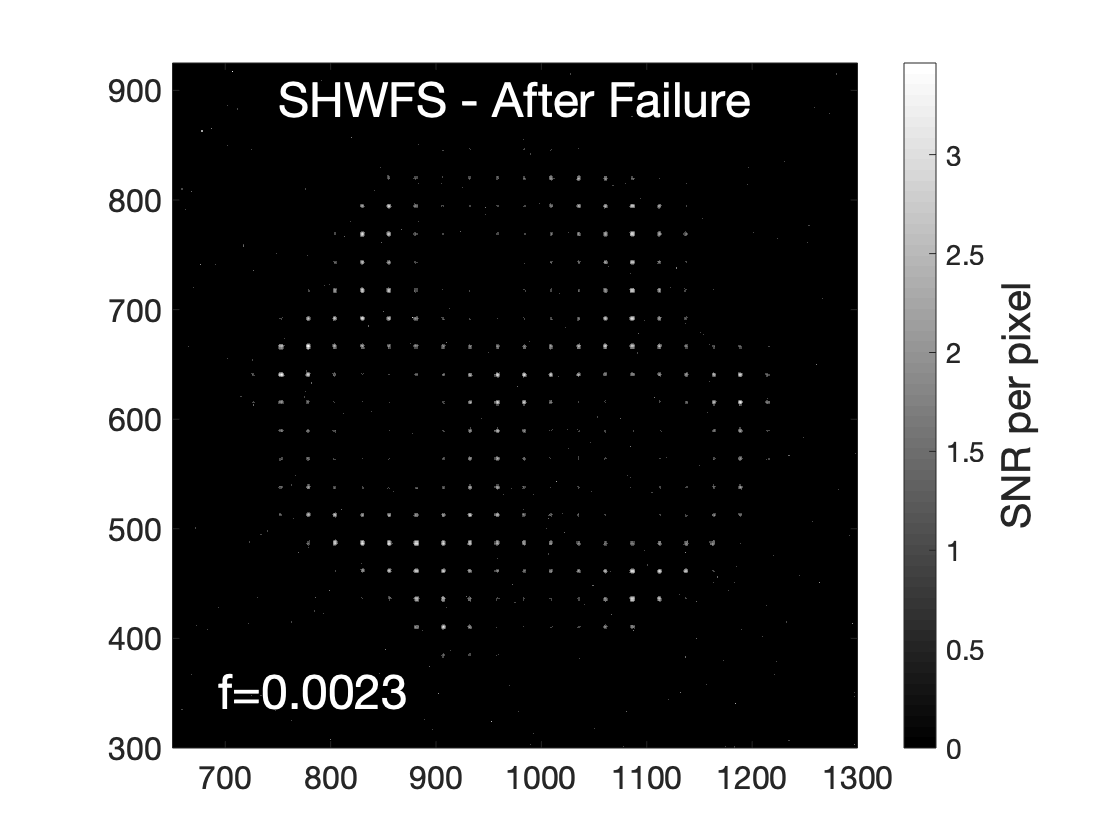}
    \includegraphics[width=0.46\textwidth]{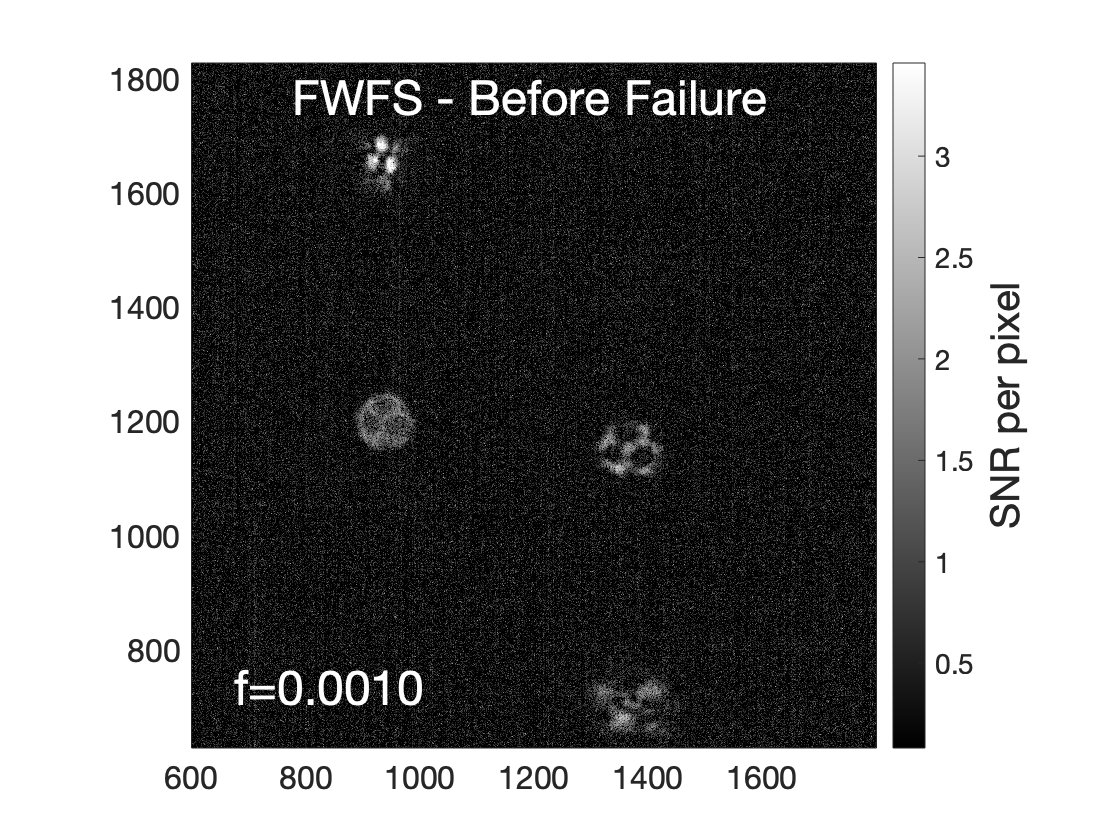}
    \includegraphics[width=0.46\textwidth]{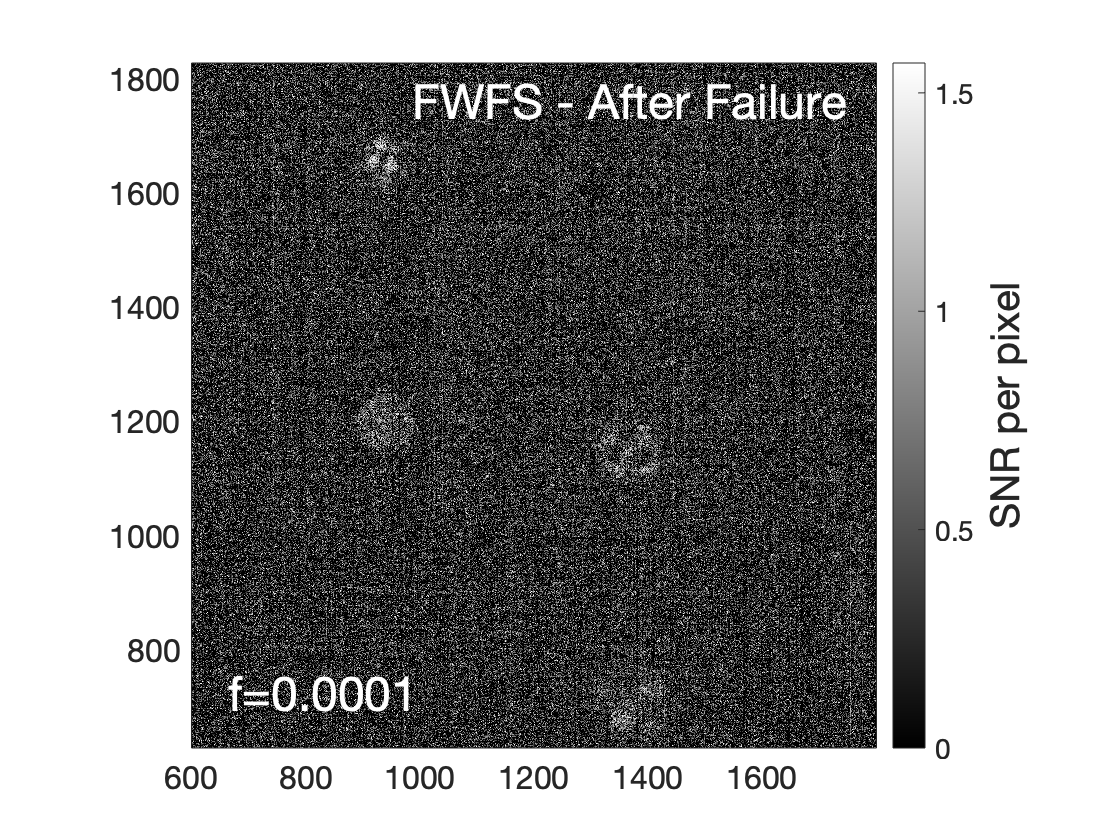}
    \caption{SNR maps recorded near the operating limit of each sensor. An intensity threshold has been applied to show measurement features.}
    \label{fig:SNR_images}
\end{figure*}

\subsubsection{Effects of Read Noise}\label{sec:readnoise}

The FWFS operates in a semi-collimated space and uses $n \approx \pi N^2$ spatial samples for sensing (four derivative beam), where $N$ is the (average) diameter of an individual diffracted beam in pixels. Read noise mixes with guide star signal and impacts the reconstruction process by introducing intensity variations that are mis-interpreted as interference. Numerical simulations show that the FWFS is sensitive to detector noise and thus \emph{requires} a low read noise camera (Potier et al., in prep.). 

Read noise of the FWFS detector is only 0.9 $e$- rms (rounded to 1 $e$- rms in the figure), whereas the SHWFS detector is 7 e- rms. For each camera, the slowest read-out modes were used to minimize read-noise. We initially found it conspicuous that the gain in sensitivity of the FWFS over the SHWFS (9$\times$) was comparable to the ratio of read noise values ($8\times$). Unfortunately, given the complexity of setting up two independent sensing channels (magnification factor, pixel size, cooling lines and available bench space), it was not practical to use the low read-noise detector for each sensor. To further study whether detector quality was the root cause of the performance difference between sensors, we: (i) artificially injected additional read noise into both the FWFS and SHWFS data sets in post-processing; and (ii) placed the more portable, higher read noise SHWFS detector into the FWFS channel.

To simulate higher degrees of camera noise, additional read noise (RN') was injected into the data by adding dark frames recorded in the lab. We found that independent dark frame realizations were statistically uncorrelated. Thus, the number of dark frames, $N_{\rm darks}$, included could be used to predictably increase the total amount of read noise (after removing a pedestal). The synthesized total read noise (RN) follows a simple quadrature relation,
\begin{equation}
    \mathrm{RN}=\sqrt{(7e-)^2 + \mathrm{RN'}^2},      
\end{equation}
where 7$e$- is the native RMS read noise of the SHWFS detector and $\mathrm{RN'}=\sqrt{N_{\rm darks}} \; 7e-$. Reconstruction results were quantified for values of $\mathrm{RN'}=[0e-, 10e-, 21e-]$, corresponding to $\mathrm{RN}=[7e-, 12e-, 22e-]$. Results were then used to estimate an equivalent $\mathrm{RN}=1e-$ curve for the SHWFS via extrapolation. 

Read noise values are indicated in the legend for each data set shown in Figure~\ref{fig:sensitivity}. Sensitivity curves with RN$=7e$- are direct measurements using the SHWFS detector. We find that adding tens of electrons of read noise to the reconstruction process has a minimal impact on the sensitivity of the SHWFS (limiting magnitude). The extrapolated 1$e$- curve is only marginally better than the 7$e$- curve. This result is perhaps not unexpected given that the SHWFS measurements occur in a focal plane. Other studies have shown that the SHWFS is capable of operating AO systems in the near-infrared with RN values of 10-20 electrons depending on source flux \citep{gendron_2003}.

In a further attempt to make an apples-to-apples comparison with the SHWFS, we also injected $\mathrm{RN'}=[2.9e-, 6.9e-]$ of read noise into the FWFS data set, corresponding to $\mathrm{RN}=[3e-, 7e-]$. Unlike the SHWFS data, we find that including additional read noise in the FWFS channel has a dramatic impact on sensitivity. Figure~\ref{fig:FWFS_RN} shows results comparing the original FWFS data taken with the Andor camera to those with additional read noise using the same wavefront reconstructor. 

Although it was not possible to place the RN$=1e$- Andor camera in the SHWFS channel, we managed to perform the reverse operation by physically removing the lenslet array from the SHWFS and inserting the RN$=7e$- SHWFS (Thorlabs) camera into the FWFS channel. Thus, as a benchmark, we also over-plot in Figure~\ref{fig:FWFS_RN} the FWFS data set that physically used the RN$=7e-$ SHWFS (Thorlabs) read noise camera.

Consistent with the above analysis, the FWFS does not perform well with 7$e$- of read noise. Indeed, the FWFS performs somewhat worse than the SHWFS when read noise is increased to from RN$=1e$- to RN$=7e$-. Increasing the read noise from RN$=1e$- to just RN$=3e$- degrades performance of the FWFS by nearly a factor of ten. We find that the higher read noise data sets become limited (at least in part) by application of the phase unwrapping algorithm, which tends to interpret sharp pixel-to-pixel intensity variations as $2\pi$ phase discontinuities. Such behavior starts to cause large variations in wavefront residuals as seen in Figure~\ref{fig:FWFS_RN} leading to non-monotonic uncertainties as a function of flux. 

Finally, we find excellent agreement between the RN$=7e$- injection read noise data set (``injected'') and RN$=7e$- measured data set (``measured''). Both sensitivity curves show similar behavior and begin to asymptote around the same flux level. It is not until the flux value falls below 1\% of the brightest sensor measurements that the curves slightly diverge. We speculate that the difference between data sets is related to response of the phase unwrapping algorithm to the different cameras used even though the read noise value was equivalent. The two RN$=7e$- data sets (injected versus measured) also were not recorded contemporaneously. Physically relocating the SHWFS (Thorlabs) camera resulted in a slight temporal drift, which can impact the structure of the wavefront residuals ($<\lambda/10$). 

The fact that the FWFS sensitivity curve shifts by a factor that is larger than the increase in read noise points towards a non-linear process. While the SHWFS is much less effected by read noise than the FWFS, its photon-noise-limited performance requires brighter beacon intensities. We find that the difference in magnitude limit (sensitivity limit) between sensors is strongly influenced by scintillation.   

\subsubsection{SNR Limits}\label{sec:snr}

Figure~\ref{fig:SNR_images} shows images of the SHWFS and FWFS channels before and after failure (defined as the first lab measurement with a software error message and/or RMS WFE exceeding 1 wave). SNR and relative flux values (again normalized to the brightest SHWFS measurements) are shown to help assess when the reconstruction procedure for each sensor breaks down. SNR per pixel was calculated by adding in quadrature photon noise plus measured background noise. Photon-noise is estimated assuming a Poisson distribution by taking the square-root of the number of photons detected in each pixel.

Upon reducing the amount of light entering each sensor, the SHWFS is able to reconstruct wavefront phase information until regions of the array experiencing destructive interference become dark. In the ``before failure'' image, spots in regions undergoing destructive interference are just discernable, while spots in regions undergoing constructive interference have SNR per pixel values of 10. In the ``after failure'' image, spots in the regions undergoing destructive interference have disappeared entirely (SNR $< 1$). All SHWFS reconstruction algorithms used are unable to operate in dim regions of the beam, causing the sensitivity curves to asymptote around the same flux level. 

The FWFS consistently reconstructs wavefront phase and amplitude information at lower flux values than the SHWFS. The bottom row of images in Figure~\ref{fig:SNR_images} shows that the FWFS is able to operate close to the fundamental photon-noise limit with SNR per pixel values near unity (when the read noise is low). The FWFS also appears to be qualitatively ``agnostic'' to the strength of scintillation, since it uses the presence of destructive interference to inform the reconstruction process. 

\section{Summary and Concluding Remarks}\label{sec:conclusions}

We have built a FWFS that offers phase diversity in the form of multiple out-of-pupil-plane images, each having different path lengths. The device uses intensity measurement planes along the optical axis to sense near-field diffraction effects. Placement of the multiple images onto a single detector minimizes size, weight, power, and cost, while eliminating differential timing latency of an equivalent multi-camera configuration.

Lab experiments were conducted in the beam control lab at the University of Notre Dame. First, a custom ``ND'' mask was fabricated to show that the FWFS can reconstruct high frequency aberrations including sharp phase transition features. Next, a phase plate was introduced into the beam to create a more continuous power spectrum of phase and amplitude aberrations. Reconstruction results using the FWFS were shown to match that of a SHWFS with equivalent spatial sampling. Measurements obtained with a scintillation index of $S=0.55$ demonstrate that the FWFS can reconstruct wavefront phase information in the presence of scintillation. Sensitivity of the FWFS and SHWFS devices were then compared by reducing the amount of light entering the experiment. The SHWFS reconstruction fails when irradiance fade patterns caused by near-field diffraction result in the under-illumination of sub-apertures. 

Despite local regions of the coherent beam disappearing entirely from destructive interference, the FWFS was able to reconstruct phase and amplitude information even when the SNR approached 2-3 per pixel. We find that the FWFS offers a factor of $\approx9\times$ improvement in sensitivity over the SHWFS at a scintillation index of $S=0.55$. The FWFS is however highly sensitive to detector noise. A very low read noise camera will be required---ideally a photon counting device---to operate an AO system in deep turbulent conditions with a FWFS. 


Stronger scintillation corresponds to larger intensity variations between regions experiencing constructive and destructive interference. Since SHWFS reconstruction algorithms fail when a threshold fraction of the sub-aperture spots disappear, the limiting magnitude of the SHWFS will degrade with progressively stronger turbulence. Thus, given the phenomenology of irradiance fade patterns, we expect that the improvement in sensitivity of the FWFS over the SHWFS depends on the scintillation index (or Rytov number). 

Overcoming the issues associated with near-field diffraction effects will facilitate AO observations that involve coherent beam propagation, laser communications, remote sensing, and other areas. Improved sensitivity offered by the FWFS over existing technologies will confer the ability to observe fainter sources or more distant sources, including situations involving strong turbulence and long path lengths. Forthcoming experiments will test the hypothesis that the FWFS can maintain a near-constant limiting magnitude with increasing scintillation index. Identical detector arrays will alleviate any remaining uncertainty in the role that read noise plays when quantifying device performance. Future research directions will explore closed-loop operation and observations of resolved sources. 

\section{Acknowledgements}\label{sec:acknowledgements}

We thank David Heemstra from the Notre Dame nano-facility for fabricating the ``ND'' masks, Matthew Engstrom from Notre Dame's Engineering and Design Core Facility (EDCF) for developing custom control software, and Scott Hampton from Notre Dame's Center for Research Computing (CRC) for assistance developing C++ code for the beam control experiment. Some of the research presented in this paper was conducted while JRC and SOL were supported by the Air Force Research Labs Summer Faculty Fellowship Program (SFFP). JHF acknowledges support from the AFRL Scholars Program. SJP acknowledges support from the NSF GRFP (Award DGE-1841556). We thank Dr. Mark Spencer for valuable technical discussions. We thank Michael Vansickle, James Dyke, and Marie Grasse for their contributions to the wavefront sensor project. 


\end{document}